\documentclass[letterpaper, 10 pt, conference]{ieeeconf}  

\IEEEoverridecommandlockouts                              

\usepackage{graphics} 
\usepackage{epsfig} 
\usepackage{mathptmx} 
\usepackage{times} 
\usepackage{amsmath} 
\usepackage{amssymb}  
\usepackage{braket}
\usepackage{hyperref}

\title{\LARGE \bf
Electronic Correlations in Multielectron Silicon Quantum Dots
}

\author{
Dylan H. Liang$^{1*}$, MengKe Feng$^{1, 2}$, Philip Y. Mai$^1$, Jesus D. Cifuentes$^{1, 2}$,
\\ 
Andrew S. Dzurak$^{1, 2}$ and Andre Saraiva$^{1, 2 \dag}$
\thanks{*haowen.liang@student.unsw.edu.au}
\thanks{\dag a.saraiva@unsw.edu.au}%
\thanks{$^{1}$School of Electrical Engineering and Telecommunications,
        University
of New South Wales, Sydney, NSW 2052, Australia}%
\thanks{$^{2}$Diraq, Sydney, New South Wales, Australia}%
}

\begin{document}

\maketitle
\thispagestyle{empty}
\pagestyle{empty}

\begin{abstract}

Silicon quantum computing has the potential to revolutionize technology with capabilities to solve real-life problems that are computationally complex or even intractable for modern computers \cite{Feynman1982-gc} by offering sufficient high quality qubits to perform complex error-corrected calculations. Silicon metal-oxide-semiconductor based quantum dots present a promising pathway for realizing practical quantum computers. To improve certain qubit properties, it is a common strategy to incorporate multiple electrons in the same dot in order to form qubits in higher confined orbital states. Theoretical modelling is an essential part of understanding the quantum behaviour of these electrons, providing a basis for validating the physical working of device models as well as providing insights into experimental data.

Hartree-Fock theory is an imperative tool for the electronic structure modelling of multi-electron quantum dots due to its ability to simulate a large number of electrons with manageable computation load. However, an efficient calculation of the self-consistent field becomes hard because dot formations in silicon are characterized by strong electron-electron interactions and conduction band valleys, besides the relatively high comparative effective mass, which add to create a behaviour dominated by repulsion between electrons rather than a well established shell structure. In this paper, we present a Hartree-Fock-based method that accounts for these complexities for the modelling of silicon quantum dots. 
With this method, we first establish the significance of including electron-electron interactions and valley degree of freedom and their implications. 
We then explore a simple case of anisotropic dots and observe the impact of anisotropy on dot formations.

\end{abstract}

\section{INTRODUCTION}

The development of quantum computers is a major physics and engineering challenge, as qubits are fragile and require the meticulous treatments to retain their quantum properties. Semiconductor quantum processors are a major contender in realizing practical quantum computation \cite{semiconductor1} \cite{semiconductor}.
Particularly, silicon MOS spin qubits confined in quantum dots are proven to be CMOS-compatible \cite{CMOScompatible}, meaning that they can leverage the existing fabrication technology to address the scalability challenge \cite{scaleup1}. 
In the last two decades, rapid development of silicon quantum processors is marked by a series of experiments that demonstrated critical features of electron-based silicon spin qubits, such as long coherence time and high fidelity two-qubit operations \cite{fidelity1, fidelity2, fidelity3}.
In principle, both electrons and holes can produce spin qubits. In this paper, we will only consider electron spin qubits.

Understanding the electronic structure of quantum dots is crucial for the effective design of functional quantum processors. 
This is not a simple task for two main reasons. 
Firstly, it is often desirable to have more than one electron confined in quantum dots to improve its performance as a qubit \cite{ross_spdf}. 
The complex interplay between electrons leads to nontrivial solutions to the total system. 
To this end, Hartree-Fock theory provides an effective method to account for electron-electron (e-e) interactions in large multi-electron systems \cite{szabo}. 
Secondly, the multiple minima or valleys in the conduction band of silicon give rise to an additional valley degree of freedom for electrons \cite{andrevalley}, which adds complexities in the shell filling of silicon quantum dots. 
We account for this by employing an atomistic tight-binding (TB) model similar to that used in ref. \cite{TBmodel} to simulate valley physics. 
By incorporating the TB model into Hartree-Fock theory, we develop a simulation model that can be applied to study the electronic structure of general semiconductor quantum dot structures.

In section \ref{method}, we outline the simulation methods. 
We first set up the mathematical problem and the Hamiltonian. 
Then, we discuss in detail how we can solve it. 
In the following section \ref{applications}, we present results from applying the Hartree-Fock method to study the electronic structure of a range of dot configurations. 
Finally, we present our conclusions and visions in section \ref{conclusion}.

\section{METHOD} \label{method}

In this section, we will discuss the implementation of the simulation method. 
Here, we focus on the algorithmic approach to carrying out the calculations and the general workflow of the method.

\subsection{Hartree-Fock Method}

In general, the solution of multi-particle systems is described by superposition of Slater determinants.
The computational complexity involved in obtaining such solutions grows exponentially with the number of particles. 
This often means the maximum number of electrons that can be solved in a reasonable time is too low for quantum processors that operate on tens or even hundreds of electrons. 
To reduce the computation load, we apply Hartree-Fock theory which approximates the ground state of the system as a single Slater determinant. 
In this way, we trade quantitative accuracy for the ability to simulate large quantum dots. In doing so, we need to carefully curate all the approximations and compare theoretical predictions to data in order to validate even the qualitative results from this method, which contains uncontrolled approximations.
We have used this method to simulate up to 20 electrons confined in an elongated silicon quantum dot, details of which can be found in ref. \cite{jellybean}.

A key consideration of Hartree-Fock theory is to treat the simultaneous interactions from all other electrons as an equivalent average potential. 
This turns an intractable many-body problem into a collection of manageable one-body problems. 
By invoking the variational principle, the best possible Hartree-Fock approximation to the ground state is obtained by minimizing the energy
\begin{equation}
    E_{\text{HF}} = \bra{\psi_{\text{AS}}}H\ket{\psi_{\text{AS}}} \label{E_hf}
\end{equation}
where $\psi_{\text{AS}}$ is a single Slater determinant.
We will construct the total system Hamiltonian $H$ in the following paragraphs.

We begin with the Hartree-Fock Hamiltonian
\begin{equation}
    H_{\text{HF}} = \sum_i H_{\text{sp}}(i) + \sum_{i<j} \nu(i,j) \label{H_hf},
\end{equation}
where the summation of $\nu(i,j)$ represents e-e interactions.
The single-particle Hamiltonian $H_{sp}$ includes the kinetic term of a single electron and the confinement potential, which can be written as
\begin{equation}
    H_{\text{sp}} = H_\text{K} + V + H_\text{Z}, \label{H_sp}
\end{equation}
where $H_\text{K}$ is the kinetic term, $V$ is the potential term and $H_\text{Z}$ is the Zeeman Hamiltonian.
We construct the Hamiltonian over a three-dimensional cuboid simulation cell with hard wall boundary conditions, divided into grids of appropriate sizes along the Cartesian directions $x$, $y$, and $z$. All results are tested for convergence over the grid finesse.

We will consider the kinetic term separately along the three directions in order to take profit of the effective mass approximation in-plane but incorporate out-of-plane valley degrees of freedom. Valleys in the lateral directions are lifted by electrostatic confinement along the vertical direction in silicon quantum dots.
So, the effects of valleys are only considered along $z$. This method is only valid for electrons pressed against the (001) interface, and will only describe adequately the energy levels below the valley states along the transversal axes $x$ and $y$.
In order to include the valley degree of freedom, we employ the second-nearest neighbor tight-binding Hamiltonian to modify the kinetic terms.
Applying the tight-binding model in the vertical direction yields
\begin{align}
    H_\text{K} = -\frac{\hslash^2}{2m_\text{t}}&(\frac{\partial^2}{\partial x^2} + \frac{\partial^2}{\partial y^2}) \notag \\
    &+ \sum_k (t_\text{1}\ket{z_{\text{k+1}}}\bra{z_{\text{k}}} + t_2\ket{z_{k+2}}\bra{z_{k}} + h.c.) \label{Hk}
\end{align}
where $t_i$ are the different hopping parameters as outlined in ref \cite{TBmodel}.
The potential term describes the confinement potential which defines the simulated quantum dot and has a different exact form depending on the problem being solved.
We will describe this in further detail in section \ref{applications} along with applications of the method.
As such, we have four spin-valley-orbitals in the Hamiltonian accounting for the charge, spin and valley degrees of freedom of electrons.

\subsection{Self-Consistent Field Procedure}
The e-e interaction term given in Eq. \ref{E_hf} involves two-electron integrals that describe the Coulomb and exchange interactions.
These integrals depend on the solution of the system, and are highly computationally expensive.
We will describe strategies for calculating them after we outline the general workflow of the overall algorithm.

We apply the self-consistent field method to iteratively minimize the total energy of the system.
The procedure begins with diagonalizing the single-particle Hamiltonian to obtain a reasonable initial guess solution.
From the guess solution, we calculate the charge density as follows,
\begin{equation}
    \rho_\text{i} = |\psi_\text{i}|^2. \label{rho}
\end{equation}
We then compute the interaction terms based on the initial guess and re-diagonalize the Hamiltonian with the interaction terms to obtain a first approximation to the solution.
Based on this first solution the mean field caused by the electrons is recalculated and the single particle solutions reobtained iteratively.
Every iteration begins with a slightly different wavefunction obtained by mixing wavefunction from previous iterations by a process described as follows,
\[\psi_{\text{new}} = (1-\delta)\psi_\text{i} + \delta\psi_{\text{i-1}},\]
where $\delta$ is a small value typically less than 10\%.

When the solution has reached self-consistency, we say the result has converged.
This is determined by monitoring the changes in total energy  $\Delta E$, wavefunction $\Delta \psi$ and exchange energy $\Delta V_{\text{ex}}$ after each iteration. We note that the exchange coupling is a small fraction of the total energy, so it needs to be separately monitored since it has a strong qualitative impact on the wavefunction (such as setting the total spin state).
The values of these quantities are compared against their corresponding convergence criteria threshold, which are numbers ranging from $10^{-4}$ to $10^{-6}$ for the differential improvement in their estimates.
Convergence is declared when the differences are below the criteria thresholds, at which point we obtain our final approximation to the ground state of the system.

As previously discussed, we require sufficiently small simulation grids to reduce errors induced by the size of the finite basis set, which typically leads to having at least 100 grid points along each of the three Cartesian directions.
This creates a Hamiltonian of size that is in the order of $100^3 = 10^6$, which is time-consuming to diagonalize.
To speed up the process, we apply a Chebyshev filtering to select a subspace of eigenvectors to map relevant eigenvalues from $(0, a)$ to $(1, \infty)$ on an even Chebyshev polynomial.
This amplifies the spacing between each eigenvalue which improves the convergence of iterative diagonalization of the Hamiltonian \cite{chebyshev}.
Details of this method can be found in ref. \cite{chebyshev2}.
On top of that, we employ the Periodic Pulay method to perform Pulay mixing (a.k.a. Direct Inversion in the Iterative Subspace (DIIS)) periodically every fixed number of iterations, as described in ref. \cite{periodicPulay}.
In our simulations, we set the period to be 6 iterations.

We now address the two-electron integrals that define e-e interactions.
Coulomb interaction is described by the Hartree potential which is defined as follows,
\begin{equation}
    V_\text{H} = \frac{e^2}{4\pi\epsilon} \int \frac{\rho(\mathbf{r^{'}})}{|\mathbf{r} - \mathbf{r^{'}}|} d^3r^{'}. \label{vH}
\end{equation}
By leveraging the efficient Fast Fourier Transform (FFT), Eq. \ref{vH} can be easily solved in the Fourier domain, in which it takes the form
\begin{equation}
    V_\text{H} = IFFT\{FFT\{\frac{1}{\mathbf{r}}\}FFT\{\rho\}\}. \label{fftvH}
\end{equation}
Similarly, exchange interaction is described by an exchange potential $V_\text{X}$, which is a full-rank $10^6 \times 10^6$ matrix.
Solving this directly will require unfeasible storage and computational power.
To realistically compute $V_\text{X}$, we employ the Adaptively Compressed Exchange method described in ref. \cite{acxLinLin} to approximate it with a low-rank version given as follows,
\begin{equation}
    V_\text{X} \approx -\sum_k \zeta_\text{k} \zeta_\text{k}^*. \label{Vx}
\end{equation}


\section{APPLICATIONS} \label{applications}

This section will present results from simulating a range of quantum dots in silicon electrostatic quantum dot devices.
The simulated quantum dot is defined by the confinement potential, which can either be analytically defined or imported from electrostatic simulations of the device model in COMSOL.
COMSOL potentials are useful for simulating quantum dots formed from realistic device models.
We can use electronic structure simulations with COMSOL potentials to compare with experimental measurements both qualitatively and quantitatively.
In this section, we are interested in studying general physical features of silicon quantum dots.
For this, we use analytic potentials that give us the freedom produce the hypothetical environments we require.

In this section, we employ simple harmonic potential landscapes on the transversal plane, which take the following form,
\begin{equation}
    V(x,y,z) = \frac{1}{2}m_{\text{x}}\omega_{\text{x}}^2x^2 + \frac{1}{2}m_{\text{y}}\omega_{\text{y}}^2y^2 - F_{\text{z}}z \label{pot}
\end{equation}
where $m_{\text{x}}=m_{\text{y}}=0.19m_0$ is the transversal effective mass of electron in silicon, $\omega_{\text{x(y)}}=\frac{E_{\text{x(y)}}}{\hslash}$ sets the confinement strength and $F_\text{z}$ is the electrical field along $z$ with a typical value of $40$meV.
If $\omega_{\text{x}} = \omega_{\text{y}}$, we say the potential is isotropic in which the confinement strength along $x$ is the same as along $y$.
In such a potential, electron wavefunctions have no preference for aligning themselves with either spatial direction, hence, will exist as some superposition state with respect to $x$- and $y$-aligned states.
We can introduce anisotropy to the confinement potential, which we denote $\alpha$.
In this paper, we define this quantity as follows,
\begin{equation}
    \alpha = \omega_{\text{y}}/\omega_{\text{x}}. \label{alpha}
\end{equation}

In the following sections, we start our discussions by exploring the two key factors affecting silicon quantum dot formation: e-e interaction and valley degree of freedom.

\begin{table}
\caption{List of relevant properties of silicon and GaAs.}
\label{properties}
\begin{center}
\begin{tabular}{|c||c|c|}
 \hline
 \textbf{Material} & Silicon & Gallium Arsenide \\ [1ex] 
 \hline
 Dielectric constant $\varepsilon_r$ & 11.7 & 12.9
 \\ 
 Transverse effective mass $m_t/m_0$ & 0.19 & 0.067 \\
 Longitudinal effective mass $m_\ell/m_0$ & 0.98 & 0.067 \\
 Typical size of quantum dot & 10nm & 40-50nm \\
 \hline
\end{tabular}
\end{center}
\end{table}

\begin{figure}
    \centering
    \includegraphics[width = 0.48\textwidth]{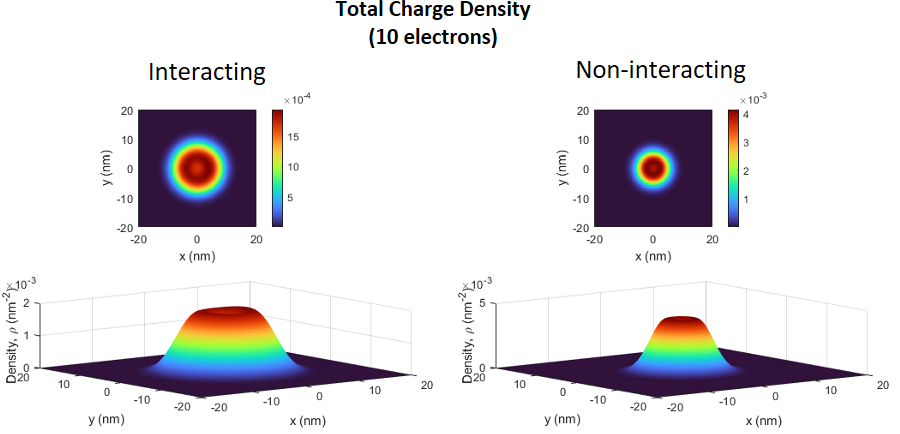}
    \caption{Total charge density for 10 electrons in the system for the interacting and noninteracting cases.}
    \label{fig:ee_chargeden}
\end{figure}

\begin{figure}
    \centering
    \includegraphics[width = 0.4\textwidth]{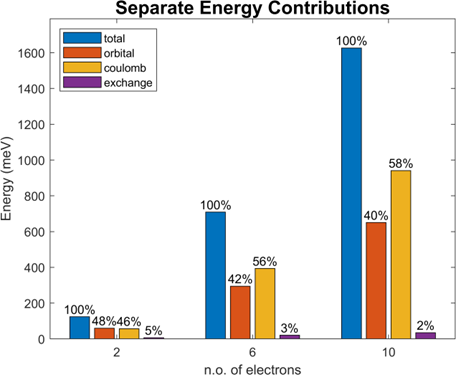}
    \caption{A breakdown of the total energy into separate energy contributions, for increasing electron numbers.}
    \label{fig:ee_segs}
\end{figure}

\subsection{Electron-Electron Interaction}

Presumably, e-e interaction has considerable contributions to the final result, especially for silicon-based processors. 
Table \ref{properties} shows a list of relevant properties of two major materials used for developing semiconductor quantum devices -- silicon and gallium arsenide. While this work is mostly concerned with silicon, this comparison serves to provide a sense of how much more important e-e interactions are in silicon than some of the other popular semiconductor platforms.

Firstly, e-e interaction scales inversely with the dielectric constant $\epsilon_r$.
The slightly lower $\epsilon_r$ in silicon than GaAs means stronger e-e interaction in principle. In a quantum dot defined at the Si/SiO$_2$ interface this difference would effectively be much larger, with the interactions enhanced by the low dielectric constant of the oxide. 
Since we do not connsider the dielectric explicitly in these calculations, this difference may be negligible in our study.
The effective mass of electrons in silicon is significantly higher than that in GaAs, even along the plane where the silicon mass is nearly five-fold smaller than out of plane.
This leads to a lower kinetic energy in silicon bound states, meaning electrons occupy a smaller region of space.
Consequently, electrons are closer to each other, so they experience stronger interactions.
Lastly, dot gates in Si-SiO$_2$ devices are typically fabricated to be only less than a quarter of the size of GaAs dots due to their closer proximity to the channel.
This further confines electrons to a tighter space, leading to stronger e-e interaction.
In conclusion, we expect e-e interaction to be much more prominent in silicon than in GaAs.
It is therefore imperative to include e-e interactions in electronic structure modelling of silicon quantum dots.
We will establish this fact in this section by examining the solutions with and without considerations of e-e interactions.

Note that e-e interaction terms carry a scaling factor of $e^2/(4\pi\epsilon)$, where the dielectric constant for silicon is $\epsilon = 11.7\epsilon_0$.
We can stablish a toy model that turns off interactions by setting $\epsilon$ to be sufficiently large such that the interaction strength is infinitesimally small.
This effectively removes the interaction terms and turn the Hamiltonian into a single-particle Hamiltonian, where all the correlations stem from the Pauli exclusion principle imposed by the Slater determinant construction of the wavefunction.

Our simulations in this section consider an isotropic harmonic potential defined in Eq. \ref{pot} where $\alpha = 1$.
With this, we first examine the total charge density of a 10-electron quantum dot for both the interacting and noninteracting scenarios, calculated according to Eq. \ref{rho}, as shown in Fig. \ref{fig:ee_chargeden}.
By comparing the charge densities side by side, it is clear that interacting electrons occupy at least double the amount of space occupied by noninteracting electrons.
This is a significant difference in size that must be considered when designing the structural dimensions and arrangement of the quantum processor. For instance, this will increase the exchange coupling between neighbouring quantum dots and impact the lever arm of the gates surrounding the dot. This phenomenon agrees with our classical intuitions that like-charges tend to stay away from each other in order to minimize Coulomb repulsion, and, hence occupying more space.

To gain a better appreciation for the significance of e-e interactions, we examine the energy of the system.
The total energy of the system given by Eq. \ref{E_hf} can be expressed in terms of separate energy contributions as the following,
\begin{equation}
    E_{\text{total}} = \sum_i E_\text{i} + \frac{1}{2}\sum_{j}V_\text{H}\sum_i \rho_\text{i} - V_\text{X} \label{segs}
\end{equation}
where the first term is the orbital energy due to the kinetic term and confinement potential, the second and third terms are the energy contributions from Coulomb and exchange interaction respectively.
We solve for 2, 6 and 10-electrons in the system and calculate the separate energy contributions with respect to the total energy as given in Fig. \ref{fig:ee_segs}.

Clearly, e-e interactions contribute to more than half of the total energy in the system, with Coulomb energy growing in percentage contribution with increasing electron numbers. We note, however, that the proportion does not increase as a linear function of the number of electrons, which indicates that the spread of the wavefunction towards regions of high potential forced by the Coulomb repulsion is equally important for the total energy compared to the direct e-e interactions.
It is not hard to imagine that these strong interactions have nontrivial implications to the formation of silicon quantum dots.
We will explore one aspect of this in a later section where we study the effects of anisotropy in confinement potentials.

\subsection{Valley Degree of Freedom}
Another crucial ingredient in simulating silicon quantum dots is the inclusion of the valley degree of freedom.
Similarly to the previous section, we explore the effects by performing simulations with and without valleys.

\begin{figure}
    \centering
    \includegraphics[width = 0.49\textwidth]{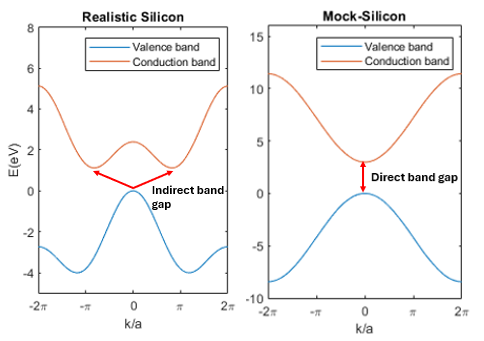}
    \caption{Dispersion relation of a realistic silicon and the mock-silicon.}
    \label{fig:dispersion}
\end{figure}

\begin{figure}
    \centering
    \includegraphics[width = 0.48\textwidth]{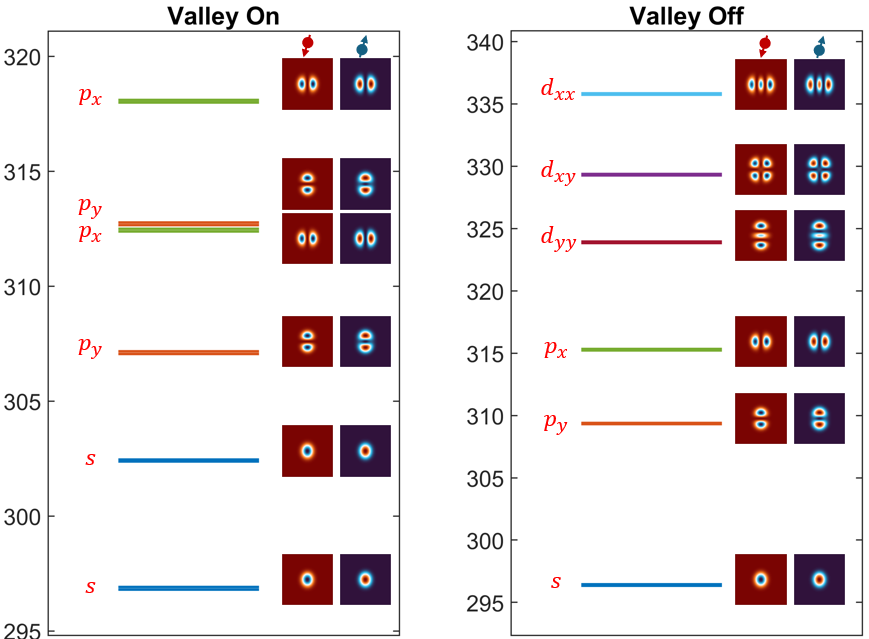}
    \caption{Energy levels and the associated occupied spin orbitals for 12 electrons in the a quantum dot in a realistic silicon (left) and the mock-silicon (right).}
    \label{fig:valley_spec}
\end{figure}

\begin{figure*}[h] 
    \centering
    \includegraphics[width = 0.95\textwidth]{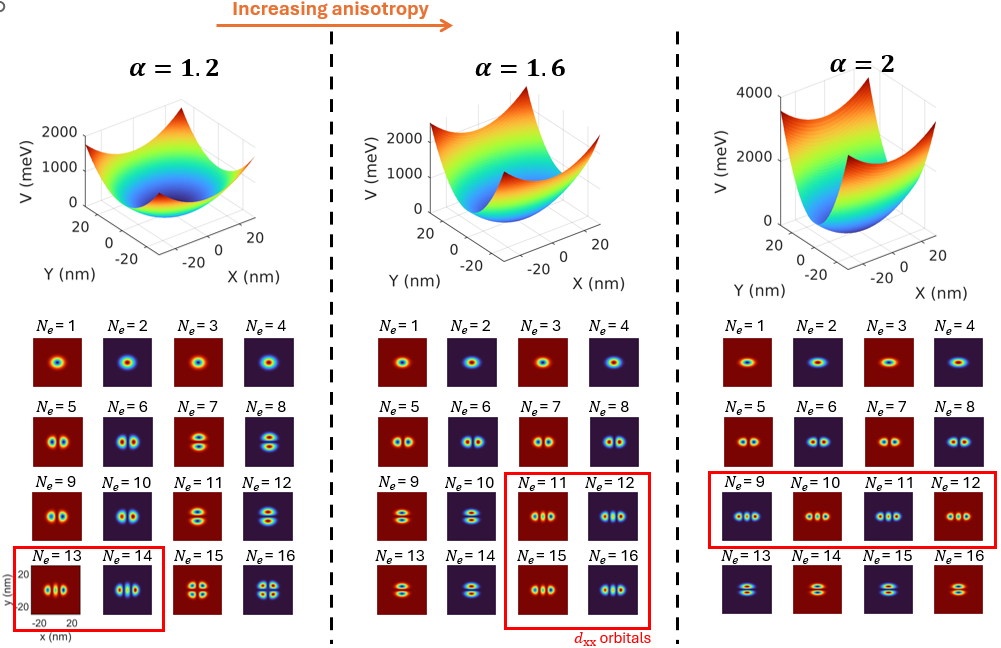}
    \caption{Confinement potentials of different anisotropy and their corresponding spin orbitals for 16 electron dots of increasing anisotropy. The red box highlights the location of $d_{\text{xx}}$-like orbitals, which move to earlier electron numbers.}
    \label{fig:aniso}
\end{figure*}

As previously discussed, valleys are accounted for by employing the tight-binding model along $z$ which yields the kinetic term as described in Eq. \ref{Hk}.
We remove the effect of valleys by defining the $z$ kinetic term as follows,
\begin{equation}
    H_{\text{k,z}} = -\frac{\hslash^2}{2m_\ell}\frac{\partial^2}{\partial z^2}
\end{equation}
where $m_\ell = 0.19m_0$ is the longitudinal effective mass of electrons.
This defines the longitudinal kinetic term in a similar fashion to the lateral ones, which effectively removes the valley degree of freedom.
We note that the Hamiltonian now describes a nonexistent material that has the parameters of a silicon without conduction band valleys.
We plot a simple dispersion relation for this `mock-silicon' and the realistic silicon in Fig. \ref{fig:dispersion}.
The presence of valleys in a realistic silicon is conveyed through the indirect band gap between the valence and conduction band.

To distinguish different orbital states, we introduce a small anisotropy $\alpha = 1.2$ to split the energy degeneracy of $x$- and $y$ orbitals.
We simulate a system of 12 electrons with and without valleys.
The orbital energy levels and their associated spin orbitals are plotted in Fig. \ref{fig:valley_spec}.
Note the orbitals diagrams with red and blue backgrounds represent spin-down and spin-up states respectively. 
In mock-silicon with valleys turned off, we see a well-defined shell filling in the order of $s$-, $p$- and $d$-like orbitals that is analogous to 3D atom shell filling dictated by Aufbau principle.
Each orbital state is occupied by a pair of electrons with anti-parallel spins separated by a small Zeeman splitting due to a $1T$ magnetic field included in the simulation.
In realistic silicon, we see each orbital state being occupied by four electrons due to the additional valley degree of freedom, resulting in 4 $s$-orbitals and 8 $p$-orbitals for 12 electrons in the system.
Evidently, valleys lead to additional energy levels in the dot which can interfere with qubit control and readout, highlighting critical importance of accounting for valleys for designing silicon quantum dot devices.

\subsection{Anisotropy}

We finalize this section with explorations of simple cases of anisotropy in silicon quantum dots.
We simulate for 16 electrons in harmonic potentials of anisotropy values of $1.2, 1.6$ and $2$.
In these simulations, we keep the confinement along $x$ constant while increasing the confinement along $y$ to increase the anisotropy according to Eq. \ref{alpha}.
The potentials and their associated spin orbital states are presented in Fig. \ref{fig:aniso}.

The spin orbitals shown are arranged in ascending order of their associated energy levels, from top left ($N_\text{e}=1$) to bottom right ($N_\text{e}=16$), which also represent the order of orbital filling.
$\alpha = 1.2$ represents the exact same scenario as the scenario in the previous section, in which the spin orbitals exhibit a somewhat standard orbital filling gradually from $s$- to $d$-like orbitals with influences from valley splitting.
Increased confinement along $y$ means electrons in $p_\text{x}$-orbitals must expend more energy to overcome Coulomb repulsion.
Eventually, it is more energetically favorable to occupy the $d_{\text{xx}}$-orbitals instead as having more nodes reduces the Coulomb repulsion. 
The red boxes in Fig. \ref{fig:aniso} highlight the gradual movement of the $d_{\text{xx}}$-orbitals towards a lower order of filling with increasing anisotropy, ultimately pushing all the $p_\text{y}$-orbitals to higher energy states.
We can similarly apply the same logic to the $d_{\text{xy}}$ states that only appear at $\alpha=1.2$, and infer that this behavior to orbitals of higher orders in the case of higher electron numbers.

\section{CONCLUSIONS AND OUTLOOK} \label{conclusion}

By simulating harmonic potentials in various hypothetical environments, we gained insights into the significance of e-e interactions and valley degree of freedom in silicon quantum dots.
As a result of strong e-e interactions, we find that anisotropy in quantum dots can have nontrivial implications on dot formation.

Much of the simulation work presented in this section can be extended for further studies. 
For one, the effects of e-e interactions and valleys established in section \ref{applications} can serve as a groundwork for comparing properties of silicon and GaAs quantum dots, which can reveal numerous useful information for the development of semiconductor quantum processors.

On the other hand, the simulation method itself can be readily applied to a range of topics outside of what was discussed in this paper, including but not limited to charge stability map, anharmonic dots and modelling structures such as nanowires.
One example of this is the application on the theoretical modelling of so-called jellybean dots for long range qubit coupling for silicon quantum processor chips as in ref. \cite{jellybean}.

\addtolength{\textheight}{-12cm}   




\section*{ACKNOWLEDGMENT}

We acknowledge support from the Australian Research Council (FL190100167 and CE170100012), the US Army Research Office (W911NF-23-10092), and the NSW Node
of the Australian National Fabrication Facility. 
The views and conclusions contained in this document are those of the authors and should not be interpreted as representing the official policies, either expressed or implied, of the Army Research Office or the US Government. 
The US Government is authorized to reproduce and distribute reprints for Government purposes notwithstanding any copyright notation herein. 
This research was undertaken with the assistance of resources and services from the National Computational Infrastructure (NCI), which is supported by the Australian Government.
D.H.L. acknowledges support from the Australian Government Department of Education through the Trailblazer Universities Program.

\bibliographystyle{ieeetr}
\bibliography{root.bib}

\end{document}